\documentstyle[aps,prl,multicol,epsf]{revtex}

\newcommand{\bleq}{\ifpreprintsty
                   \else
                   \end{multicols}\vspace*{-3.5ex}{\tiny
                   \noindent\begin{tabular}[t]{c|}
                   \parbox{0.493\hsize}{~} \\ \hline \end{tabular}}
                   \fi}
\newcommand{\eleq}{\ifpreprintsty
                   \else
                   {\tiny\hspace*{\fill}\begin{tabular}[t]{|c}\hline
                    \parbox{0.49\hsize}{~} \\
                    \end{tabular}}\vspace*{-2.5ex}\begin{multicols}{2}
                    \fi}
\newcommand{\bcols}{\ifpreprintsty\else\begin{multicols}{2}\fi}
\newcommand{\ecols}{\ifpreprintsty\else\end{multicols}\fi}

\begin{document}
\bibliographystyle{prsty}
\title{Counting Statistics of an Adiabatic Pump}
  
\draft

\author{Anton Andreev$^{1}$, and Alex Kamenev$^{2}$}
\address{$^{1}$ Department of Physics, CB 390,
 University of Colorado, Boulder, CO 80303, USA.\\
 $^{2}$ Department of Physics, Technion, Haifa 32000, Israel.
  \\
  {}~{\rm (\today)}~
  \medskip \\
  \parbox{14cm} 
    {\rm We consider quantum statistics of charge transmitted through a mesoscopic
      device in the adiabatic pumping process. 
      A general formula for the distribution  
      function of the transmitted charge in terms of the time--dependent
      S--matrix  is obtained. It is applied to a
      few simple examples of the pumping cycles. 
      We  identify coherent pumping strategies which lead to the binomial 
charge distribution and minimize the fluctuations of the pumped current. 
Conditions for the ideal noiseless quantized pump are discussed.  
     }\bigskip \\ }

\maketitle

\bcols

Adiabatic charge pumping has attracted 
considerable theoretical and experimental interest. 
It occurs when the Hamiltonian of the system is changed periodically with 
time. At the end of the pumping cycle a finite charge may be transmitted
through the system.  Such a charge transfer takes place even in the
absence of any dc voltage applied to the system. The idea is originally due
to Thouless\cite{Thouless83}, who showed that in certain one-dimensional
systems the transmitted charge is quantized in the adiabatic limit. 
This mechanism has been proposed for making electric current standards~\cite{Niu}. 

Recently research has focussed on adiabatic pumping through mesoscopic
devices~\cite{Beal-Monod,Brouwer98,Shutenko,Marcus,Aleiner98,Kouwenhoven}.
Most of the  work concentrated on the average pumping
current and its mesoscopic
fluctuations~\cite{Beal-Monod,Brouwer98,Shutenko,Marcus}. 
The general expression for the average transmitted charge
was derived by Brouwer~\cite{Brouwer98} in terms of the time--dependent
$S$-matrix based on the formalism of Ref.~\cite{Buttiker}. 
The {\em average} charge transmitted during a cycle 
depends on the system's conductance and is not in general
quantized~\cite{Aleiner98,Shutenko}. The absence of quantization 
is related to the fact that the true Thouless adiabatic conditions 
\cite{Thouless83} can never be achieved for a system with infinite leads
and hence a vanishing excitation gap. The possibility to create
electron-hole pairs with arbitrarily small energy leads to dissipation and
violates the exact quantization~\cite{Shutenko}.  
Taken together with the discrete nature of charge carriers this fact 
suggests the presence of thermal and  shot noise in the adiabatic 
pumps. A better understanding of the noise characteristics of such devices 
is  certainly necessary if the prospect of their use as current standards 
\cite{Kouwenhoven} is to be taken seriously.

In this Letter we address the problem of charge counting statistics 
in an adiabatic pump. We have greatly benefited from the extensive work
on the current statistics of voltage-biased resistors by Levitov
et. al.~\cite{Levitov,coherent,Ivanov83}. 

We consider  a scatterer connected with 
the left and right leads, each having $n$ transverse channels~\cite{foot1}. 
Such a scatterer is characterized by 
a $ 2n\times  2n$ unitary scattering matrix $S$, 
which can be written 
in the $n\times n$ block form
\begin{equation} 
S=\left(   \begin{array}{ll} 
r & t' \\
t & r' 
\end{array} \right)\, .
                                                             \label{Smat}
\end{equation}
Here $r$ ($r'$) and $t$ ($t'$) are left (right) reflection and transmission 
matrices correspondingly. 
 We shall assume hereafter that the $S$-matrix 
is a periodic function of time $\tau$, with the period $\tau_0$, 
$S(\tau+\tau_0) =S(\tau)$. 
This time--dependence is supposed to be adiabatic,   
i.e. $\tau_0$ is 
much greater than the Wigner delay time of the device. We restrict our
attention to open systems, i.e. those with the dwell time shorter than the inverse
mean level spacing in the device. Under these conditions one  may  neglect
both the energy relaxation  and the Coulomb
blockade effects inside the device.

The transmission and reflection matrices in Eq.~(\ref{Smat})
can be simultaneously diagonalized by the {\em  block-diagonal} 
unitary matrices 
$U(\tau)$ and $V(\tau)$ (see for example Ref.~\cite{Stone95})
\begin{equation} 
S(\tau) = U(\tau) {\tilde S}(\tau) V^{\dagger}(\tau)\, .
                                                             \label{tdep}
\end{equation}
Here 
${\tilde S}(\tau)$ is a matrix of the form Eq.~(\ref{Smat}) with real 
diagonal reflection and transmission blocks. The eigenvalues of the latter
play the role of the instantaneous  
reflection and transmission amplitudes in the channels. 
The ambiguity in the definition of matrices $U(\tau)$ and $V(\tau)$ does
not affect our final results.

In each cycle a  number of electrons $Q$ may pass through the scattering
region in a  direction which depends on the detailed form of $S(\tau)$.
The average charge transmitted in $N$ pumping cycles 
was recently given by Brouwer~\cite{Brouwer98}  
\begin{equation} 
\langle Q \rangle =
{1\over 2i} \int\limits_0^{N\tau_0}\!\! {d \tau\over 2\pi}\,   
\mbox{Tr}    \left\{
{\partial S\over\partial \tau} S^{\dagger} \sigma_3 
\right\}\, ,
                                                             \label{aver}
\end{equation}
where charge $Q$ is measured in  units of the electron charge. 
We have used the representation where  the transmitted charge is written
as  half the sum of that through the left and through the right leads 
\cite{foot2}, 
thus $\sigma_3={\rm diag}\{1,-1\}$ is a Pauli matrix with  the $n\times
n$ block structure. 
Both quantum and thermal noise lead to fluctuations in the transmitted charge 
$Q$. As a result the charge transmitted in one cycle can be described by 
a certain probability distribution. 

Here we obtain the distribution function of transmitted charge for 
an adiabatic pump characterized by a periodic 
time-dependent $S$-matrix, $S(\tau)$.   
More precisely,  we calculate the probability  $P_N(Q)$ to 
transmit the charge $Q$ upon completion of $N$ pumping cycles. 
It is convenient to formulate the results in
terms of the generating function $F_N(\lambda)$ of the 
moments of the transmitted charge defined as~\cite{Levitov}  
\begin{equation} 
F_N(\lambda)= \langle e^{i Q\lambda} \rangle  = \int d Q
 P_N(Q) e^{i Q\lambda}\, .
                                                             \label{gfun}
\end{equation}
For the $S$-matrix of the form Eq.~(\ref{tdep}) we obtain
\begin{equation} 
F_N(\lambda)\! = \!  e^{i\hat N\lambda}
\mbox{det}\! \Big[ 1\! +\! {\tilde S}_{\lambda}(\tau) {\tilde n}
(\tau\!, \!\tau') 
({\tilde S}^{\dagger}_{-\lambda}(\tau')\! -\! {\tilde
  S}^{\dagger}_{\lambda}(\tau'))\! \Big]\, ,
                                                             \label{main}
\end{equation} 
where 
\begin{equation} 
{\tilde S}_{\lambda}(\tau) \equiv 
\exp\{-i\sigma_3\lambda/4\} {\tilde S}(\tau) \exp\{i\sigma_3\lambda/4\}\, .
                                                             \label{Slamb}
\end{equation} 
The  matrix $ {\tilde n}(\tau,\tau')$ is defined as 
\begin{equation}
  \label{eq:ntilde}
  {\tilde n}(\tau,\tau')=V^\dagger (\tau) \hat{n}(\tau -\tau') V(\tau').
\end{equation}
Here the diagonal matrix $ {\hat n}(\tau-\tau')$ is the time Fourier
transform of  
$\hat n(\epsilon) = \mbox{diag}\{n_L(\epsilon), n_R(\epsilon)\}$, where 
$n_{L(R)}(\epsilon)$ is the energy distribution function of the left (right) 
lead. The operator in  the determinant in  Eq.~(\ref{main}) should be 
understood as an operator in the time space as well as a matrix 
in the space of channels.   
Finally,  the {\em integer}  number $\hat N$  defined as 
\begin{equation} 
\hat N = {1\over 2i} \int\limits_0^{N\tau_0}\!\! {d \tau\over 2\pi}\,  
\mbox{Tr} \left\{ 
 U^{\dagger} {\partial U\over\partial \tau}\sigma_3  - 
 V^{\dagger} {\partial V\over\partial \tau}\sigma_3  
\right\} \,  ,
                                                             \label{hatN}
\end{equation}
is the contribution to the generating function arising from the chiral
anomaly (see below). Equation (\ref{main}) is the main result of this
Letter.  Before presenting the   
derivation, we shall illustrate its applications on a few examples.

Expanding $\ln F_N(\lambda)$ to the first power of $i\lambda$ one finds for 
the average transmitted charge, 
$\langle Q \rangle \equiv\sum_Q QP_N(Q)$, 
\begin{equation} 
\langle Q \rangle =\!\!
\int\limits_0^{N\tau_0}\!\! {d \tau\over 4\pi}\, 
\mbox{Tr} \big\{{\tilde S}(\tau) {\tilde n}(\tau, \tau')
[\sigma_3,{\tilde S}^{\dagger}(\tau')] \big\} \Big|_{\tau' \to \tau} \!\!\!\!
+ \hat N\, . 
                                                             \label{aver1}
\end{equation}
In the absence of an external voltage ${\tilde  n}(\tau,\tau')\to
i/[2\pi(\tau-\tau'+i\eta)] + i/(2\pi) V^\dagger(\tau) \partial
V(\tau)/\partial \tau$ for $\tau' \to \tau$. Expanding $\tilde S^{\dagger}(\tau')$
to the first 
power in $\tau-\tau'$ and  using Eq.~(\ref{hatN}) one obtains 
Brouwer's result~\cite{Brouwer98}, Eq.~(\ref{aver}).

A particularly instructive example is given by a single channel 
case ($n=1$) with the $S$-matrix depending on time as    
\begin{equation} 
S(\tau) = e^{i\sigma_3 \theta(\tau)/2}\tilde S e^{i\sigma_3
  \theta(\tau)/2}\, , 
                                                             \label{examp}
\end{equation} 
where $\theta(\tau+\tau_0)-\theta(\tau) =2\pi$, and $\tilde S$ is
time-independent.  This form of the $S$-matrix is 
equivalent to the phase winding of the reflection amplitudes,
$r\to r \exp\{ i \theta(\tau)\}$ ($r'\to r' \exp\{ -i \theta(\tau)\}$). 
The contribution from the chiral anomaly in Eq.~(\ref{hatN}) coincides 
with the number of cycles, $\hat N = N$.  
We concentrate first on a particularly simple time--dependence,
$\theta(\tau) = 2\pi \tau/\tau_0$. 
In this  case the determinant in Eq.~(\ref{main}) may be 
easily calculated in the Fourier basis leading to
\begin{eqnarray} 
F_N(\lambda) &=&  
\prod\limits_k 
\Big[ 
1+n_L(\epsilon_{k+N})(1-n_R(\epsilon_k) )(e^{i\lambda}-1)|t|^2 
                                          \nonumber \\
&+& (1-n_L(\epsilon_{k+N}) ) n_R(\epsilon_k) (e^{-i\lambda}-1)|t|^2
\Big]  
e^{i N \lambda }\, ,
                                                             \label{single}
\end{eqnarray}   
where $\epsilon_k=\pi(2 k+1)/(N\tau_0)$ are fermionic frequencies. 
At small temperature, $T\ll  1/\tau_0, V$,  
this expression simplifies substantially, leading to 
\begin{equation} 
F_N(\lambda)= 
(e^{i\lambda}|r|^2 +|t|^2)^N 
(|r|^2 + e^{-i\lambda}|t|^2)^{N \tau_0 V\over 2\pi} \, ,
                                                             \label{binom}
\end{equation}   
where $V$ is a voltage applied between the left and right leads, which 
is chosen to be such that $N\tau_0 V/(2\pi)$ is an integer \cite{Levitov}. 
In the absence of external voltage Eqs.~(\ref{gfun}) and (\ref{binom}) lead to 
the {\em binomial} distribution function of the charge transmitted through the 
adiabatic pump
\begin{equation} 
P_N(Q)= C^Q_N\, |r|^{2Q}\, (1-|r|^2)^{N-Q}\, .
                                                             \label{binom1}
\end{equation}  
Note that only the integer values of the transmitted charge have non--zero 
probability to be detected. 
The physical meaning of this expression is that each pumping cycle is 
associated with an attempt to transfer one electron. The success probability 
of such an attempt is given by the reflection probability, $|r|^2$, 
whereas the probability of failure is 
$1- |r|^2$. 
The above statistics should be 
compared with the case of the dc voltage applied across the scattering
region  (no pumping). This case 
may be obtained from Eq.~(\ref{binom}) in the limit $N\to 0$, whereas 
$N\tau_0 V/(2\pi) \to \tilde N$ -- integer. We immediately recover the
familiar  results \cite{Levitov}  
\begin{equation} 
P_{\tilde N}(Q) = C^Q_{\tilde N}\, |t|^{2Q}\, (1-|t|^2)^{\tilde N-Q}\, .
                                                             \label{binom2}
\end{equation}  
The distribution is also binomial, however the probability of success
is given by the transmission probability, $|t|^2$. The second 
cumulant of the transmitted charge coincide for the adiabatic pump 
and applied dc voltage and is given by 
\begin{equation} 
\langle\langle Q^2 \rangle\rangle =  |t|^{2} (1-|t|^2) N \, ,
                                                             \label{noise}
\end{equation}   
leading to the maximal noise power in both cases at $|t|^2=|r|^2=1/2$.

The binomial distribution, Eq.~(\ref{binom1}) was derived above for the 
simplest time dependence of the form 
$r(\tau) = r\exp\{2\pi i \tau/\tau_0\}$.
The same result  may be shown to be valid for a more general class of 
pumping strategies, which we call the coherent pumping, 
following the terminology of Ref.~\cite{Levitov,coherent}:
\begin{equation} 
e^{i\theta(\tau)} = 
\frac{ e^{2\pi i \tau/\tau_0} - z} { 1 - z^* e^{2\pi i \tau /\tau_0} }\, ,
                                                             \label{coherent}
\end{equation}   
where $z$ is a complex number with $|z| < 1$. 
Indeed, at zero temperature $n(\epsilon_k)=\theta(-2k-1)$ 
has infinitely degenerate eigenvalues. One can therefore 
diagonalize the operator in Eq.~(\ref{main}), in the $z$-dependent 
basis which does not mix positive and negative frequencies. 
The value of the determinant does not depend on $z$ and one recovers the
binomial distribution (\ref{binom1}).
Such coherent pumping strategy guarantees the minimal possible value of 
the noise in the transmitted charge. The general expression
for the noise of the adiabatic pump may be derived by
expanding $\ln F_N(\lambda)$,  Eq.~(\ref{main}),  to the second order in
$i\lambda$ and is given by
\begin{equation}  
\langle\langle Q^2\rangle \rangle =  \int\!\!\!\int\limits_0^{N\tau_0}\!\! 
{d\tau d\tau' \over (4N\tau_0)^2 }
\frac{\mbox{Tr} \left\{ 
1- (S^{\dagger}\sigma_3 S)_\tau (S^{\dagger}\sigma_3 S)_{\tau'} \right\} }
{\sin^2 [\pi(\tau-\tau')/(N\tau_0)] }\, .
                                                             \label{noise1}
\end{equation}  
Substituting the $S$--matrix of the form Eq.~(\ref{examp}) and minimizing 
the charge fluctuations, $\langle\langle Q^2\rangle \rangle$,  
with respect to $e^{i\theta(t)}$~\cite{Korshunov87}, 
one finds that the coherent pumping, Eqs.~(\ref{examp}), (\ref{coherent}), 
lead to  the minimal possible noise.  The value of this minimal 
noise is given by Eq.~(\ref{noise}).   Note also that the second moment, 
Eq.~(\ref{noise1}) (as well as the first one, Eq.~(\ref{aver}), and the
higher ones) may be expressed through the $S$--matrix only, rather than
through auxiliary matrices defined in Eq.~(\ref{tdep}).

Next we consider a $2 \times 2$ scattering matrix with real 
reflection and transmission amplitudes given by $r= -r'=\cos (2\pi \tau
/\tau_0)$ and $t=t'=\sin (2\pi \tau /\tau_0)$ respectively. 
In this case $U(\tau)=V(\tau)=1$ leading to $\hat N=0$, 
and $\tilde n$ is the equilibrium distribution function. 
One can show that $F_N(\lambda)$ in Eq.~(\ref{main}) is an even function
of $\lambda$, and is therefore real. 
As a result one may employ the method of
Ref.~\cite{Ivanov83} to compute the determinant of 
the operator in Eq.~(\ref{main}): one multiplies this operator 
by its Hermitian conjugate and takes the square root.  
The resulting operator may be written as $ 1 + (1-\tilde n)\tilde
S_\lambda^\dagger \tilde S_{-\lambda} \tilde n +  \tilde n
S_{-\lambda}^\dagger  \tilde S_{\lambda}(1-\tilde n)$. 
In the energy representation it has a finite number of the
off-diagonal matrix elements and its determinant can be straightforwardly 
evaluated. This way  one obtains
\begin{equation}
  \label{eq:nonbin}
  F_N(\lambda)=\left (\frac{1+\cos \lambda }{2}\right)^N. 
\end{equation}
Therefore such pump is a realization of the ``random walk motion'' for charge.
Indeed, there are three possible values of the transmitted charge in each 
cycle:   $Q=0$ with the probability $1/2$, and $Q=\pm 1$ with
the probability $1/4$ each.

We note that the logarithmic derivative $iV^\dagger(\tau)\partial
V(\tau)/\partial \tau$ is analogous to the instantaneous
matrix of ``voltages'' applied to the {\em    incoming} 
channels. The integral of this quantity can be interpreted as the number
of transmission attempts~\cite{Levitov}. The different outcomes of
such attempts lead to the noise of the pumping current. In general the
probability distribution of the transmitted charge is not binomial.  
If the ``voltage'' matrix can not be
diagonalized simultaneously with the reflection and transmission matrices
then the distribution function of the transmitted charge does not
factorize into binomial distributions of elementary transmission
processes.

In contrast, the matrix $U(\tau)$ corresponds to the {\em outgoing}
channels and enters the final expression (\ref{main}) only through
the chiral anomaly term (\ref{hatN}) and therefore contributes to the average
current but not to the noise. For example, the pumping cycle of the form
$S(\tau) =U(\tau) \tilde S$ at zero temperature would produce a noiseless
quantized pumping current.

We turn now to the derivation of Eq.~(\ref{main}). To this end we model
the leads by a $2n$-component vector of chiral incoming fermions 
$(\psi_{L} (x,\tau), \psi_{R}(x,\tau))$ 
and $2n$-component vector of chiral outgoing fermions 
$(\xi_L (x,\tau), \xi_R(x,\tau))$.
The action for e.g. left lead is written as 
\begin{equation}  
S_L\! =\!\! \int\limits_{\cal C}\!\!\! d\tau\!\! \!
\int\limits_{-\infty}^0\!\!\! dx\,
\bar \psi_L(\partial_t \! + \! \hat v_L \partial_x)\psi_L \! + \!
\bar \xi_L(\partial_t  \! - \! \hat v_L \partial_x)\xi_L\, , 
                                                             \label{action}
\end{equation}  
where $\hat v_L$ is a diagonal $n\times n$ matrix of the left lead channel 
velocities. In this expression 
the time integral runs along the Keldysh contour, ${\cal C}$, 
from $\tau=0$ to $\tau=N\tau_0$ and then back to $\tau=0$. 
The right lead is described by the similar action with 
the space integral running from $x=0$ to $x=+\infty$, and the velocity matrix 
$\hat v_R$. Finally the incoming and outgoing channels at $x=0$ are related 
by the  time--dependent $S$-matrix operator
\begin{equation}  
\xi(0,\tau) =  \hat v ^{1/2}S(\tau) \hat v ^{-1/2} \psi(0,\tau) \, .
                                                           \label{bound}
\end{equation}

The current operator has a form  
$I = ( I_L +  I_R)/2$, where 
\begin{equation}  
 I_L(\tau)\! = \! 
\big[ \bar\psi_L(0^-,\tau)\hat v_L \psi_L(0^-,\tau) \! - \!
\bar\xi_L(0^-,\tau)\hat v_L \xi_L(0^-,\tau) \big]\, . 
                                                           \label{current}
\end{equation}  
The operator of the charge transmitted in $N$ cycles is given by 
$Q = \int\limits_0^{N\tau_0} d\tau I(\tau)$. 
Finally, the  generating function  may be written as
\begin{equation}  
F_N(\lambda) =\! \int\!\! D[\psi,\xi] \, 
e^{-S_L-S_R + {i\over 2}\int\limits_{\cal C} d\tau \hat \lambda(\tau) 
I(\tau) }\, ,
                                                         \label{F}
\end{equation}  
where $\hat \lambda(\tau)$ is equal to $\lambda$ on the forward and 
$-\lambda$ on the backward part of the Keldysh contour.  The fermion 
fields in this integral obey the boundary condition, Eq.~(\ref{bound}).
One has to specify the initial, $\tau =0$, density matrix, which 
implicitly defines the Green functions. 
We fix the occupation numbers in the incoming 
channels of the left and right leads to be $n_L(\epsilon)$ and 
$n_R (\epsilon)$ correspondingly, whereas the outgoing channels are supposed 
to be initially empty in accord with the scattering setup.   
  
The subsequent calculations amount to the evaluation of the Gaussian integral 
in Eq.~(\ref{F}). To this end we first make the  {\em chiral} gauge transformation
of the fermionic fields:
$\psi(x,\tau)\to V(\tau)\psi(x,\tau)$ and 
$\xi(x,\tau)\to U(t)\xi(x,\tau)$. 
As a result, the boundary condition for the new fermions  contains the 
$\tilde S (\tau)$ matrix only and the action acquires an additional 
time--dependent (matrix) chemical potential term
\begin{equation} 
\delta S \! = \! \int\limits_{\cal C}\!\! d\tau \! \int\!\! dx \,
\bar \psi \left[ V^{\dagger} {\partial V\over\partial \tau} \right] \psi + 
\bar \xi  \left[ U^{\dagger} {\partial U\over\partial \tau} \right] \xi\,  .
                                                             \label{gauge}
\end{equation}
Such potential term results in the redefinition of the density matrix
according to Eq.~(\ref{eq:ntilde}). 
Importantly, upon the chiral gauge transformation the expression for the
current acquires an extra term $I \to I +  1/(4\pi i){\rm Tr} 
\left\{ \partial
 U^\dagger (\tau)\sigma_3 U(\tau)/\partial \tau-\partial
V^\dagger (\tau)\sigma_3V(\tau)/\partial \tau  \right\}$
arising from the chiral anomaly~\cite{Zinn-Justin}.

Since the source field, $\hat \lambda(\tau)$, is a constant on the both  
branches of 
the Keldysh contour, one may eliminate 
the $\int \hat \lambda I$ term 
 from the action 
 by the time--independent gauge transformation \cite{Levitov}, e.g. 
$\psi_L \to e^{i\theta(x-0^-)\lambda/2} \psi_L$ on the forward branch 
of the contour 
and  
$\psi_L \to e^{-i\theta(x-0^-)\lambda/2} \psi_L$ on the backward branch. 
Such  transformation leads to the change in the phase of the forward
scattering amplitude and can be taken into account by a redefinition 
of the $\tilde S$--matrix  in the boundary condition, 
Eq.~(\ref{bound}), $\tilde S\to \tilde S_{\pm \lambda}$ on the forward 
(backward) branches, with $\tilde S_{ \lambda}$  defined
in Eq.~(\ref{Slamb}).

The subsequent steps are straightforward. One integrates out 
all degrees of freedom except for those which reside directly at the 
scatterer, $x=0$. Using the boundary condition, 
Eq.~(\ref{bound}), with $\tilde S_{ \pm \lambda}$--matrix, 
one eliminates the incoming degrees of freedom, $\psi(x=0, \tau)$. 
The remaining Gaussian integral over the outgoing 
fermions, $\xi(x=0, \tau)$, can be straightforwardly       
evaluated, resulting in the determinant written in Eq.~(\ref{main}). 
The remaining term, $\exp\{i\hat N \lambda\}$, is  the contribution from 
the chiral anomaly, as explained above.   

To conclude, we have derived a general expression for the counting 
statistics of the charge transmitted through  a system described by a
time--dependent $S$--matrix. The only limitations of our result are the
requirements of adiabaticity and the absence of inelastic processes in the
scattering region. 
 The absolute minimum of the noise power may be achieved
by the coherent pumping strategy, in which case the charge distribution
is given by the product of binomial distributions. 
We point out the major role played by the chiral anomaly contribution
to the average transmitted charge. Such anomalous term did not arise in
the context of voltage--biased systems  \cite{Levitov}, but is extremely 
important for the adiabatic pumping setup.

It is our pleasure to acknowledge helpful discussions with I.~Aleiner,
Y.~Avron, B.~Spivak and L.~Sadun. 
We appreciate the warm hospitality of the Norwegian Centre for
Advanced Studies, where part of this work was performed.  This work was
partly supported through the grants BSF-9800338  and DMR-9984002. 
A.~A. is an A.~P.~Sloan and Packard Fellow.

\ecols

\begin{thebibliography}{10}
\bibitem{foot1} One can
  always add some ficticious non-conducting  channels to one of the leads
  to make the number of channels in each lead equal.
\bibitem{foot2}
To have the average charge passing in and out the system equal,
the $S$--matrix should satisfy 
$\int_0^{\tau_0} d\tau \mbox{Tr} \{S^{\dagger} \partial S/\partial \tau \} = 0$.
\end{thebibliography}
\end{document}